\documentclass[pre,twocolumn,showpacs,floatfix]{revtex4}
\usepackage{graphicx}
\usepackage{amsmath}
\usepackage{amsfonts}
\usepackage{amssymb}
\usepackage{hyperref}
\usepackage{bm}
\hypersetup{bookmarks,colorlinks,citecolor=red,filecolor=blue,urlcolor=blue}
\DeclareGraphicsExtensions{.pdf,.jpg,.png}

\providecommand{\e}[1]{\ensuremath{\times 10^{#1}}}

\begin{document}
\title{Frequency Precision of Two-Dimensional Lattices of Coupled Oscillators with Spiral Patterns}
\author{John-Mark A. Allen}
\affiliation{Corpus Christi College, University of Cambridge, Cambridge, CB2 1RH, UK}
\author{M. C. Cross}
\affiliation{Department of Physics, California Institute of Technology, Pasadena CA 91125, USA}

\date{\today}

\begin{abstract}
Two-dimensional lattices of $N$ synchronized oscillators with reactive coupling are considered as high-precision frequency sources in the case where a spiral pattern is formed. The improvement of the frequency precision is shown to be independent of $N$ for large $N$,
unlike the case of purely dissipative coupling where the improvement is proportional to $N$,
but instead depends on just those oscillators in the core of the spiral that acts as the source region of the waves. Our conclusions are based on numerical simulations of up to $N=29,929$ oscillators, and analytic results for a continuum approximation to the lattice in an infinite system. We derive an expression for the dependence of the frequency precision on the reactive component of the coupling constant, depending on a single parameter given by fitting the frequency of the spiral waves to the numerical simulations.
\end{abstract}

\maketitle

\section{Introduction} \label{sec:introduction}

Frequency sources find uses in applications such as timekeeping, communications, and sensors. Crucial for all applications is the level of frequency precision, and increasing frequency precision is a necessity for improving these technologies. No practical oscillator is ever completely isolated, but subject to stochastic noise: it is this noise that reduces the frequency precision. The frequency precision of oscillators subject to noise is also important in biological systems \citep{Winfree01}.

Traditional frequency devices (e.g.\ quartz crystals) have fundamental size limitations: it is a current engineering challenge to increase frequency precision with smaller devices. An alternative frequency technology is the nano-scale mechanical oscillator \citep{Ekinci05,Kenig12}, which is much smaller than current frequency sources but also much less precise. It is a well known property of oscillators that when they are coupled they tend to \emph{synchronize} \citep{Pikovsky01} to a common frequency. It has been suggested that by coupling $N$ oscillators the resulting array, treated as a single oscillator, may have enhanced frequency  precision \citep{Winfree01,Chang97,Needleman01}.
Such a method exploits the fact that the noise sources on constituent oscillators are independent. Therefore, if one were to average over independent oscillators for a long time, the effect of the noise would reduce by $N^{-1}$. Simply averaging the signal from independent oscillators is not viable since there will be small differences in the natural frequencies due to manufacturing imperfections or natural variability, and so the signal itself would also vanish. Synchronization prevents this, but the collective dynamics of the system becomes non-trivial, and the simple argument for noise reduction does not in general apply.

When coupled, the dynamics of an oscillator become a function of the difference in phase to each oscillator with which it is coupled.  Coupling is \emph{dissipative} if it is an odd function of phase difference, and \emph{reactive} if an even function \citep{Topaj02}: a physical realization may include both components. Reactive coupling causes differences in phase to propagate through the network of oscillators, and the synchronized state is often one of waves propagating away from one or more sources in the lattice \citep{Sakaguchi88}. In this situation, it might be anticipated that the frequency precision of the resulting state should depend on properties of these source regions, rather than the lattice as a whole. Masuda \emph{et al.}\ \citep{Masuda10} demonstrated this for a lattice of identical oscillators except for a small region of higher frequency ``pacemakers'' they added to act as the source. They showed that  the improvement in frequency precision is not the na\"{i}ve $N^{-1}$ in the case of nonzero reactive coupling. One of us \citep{Cross09} extended this to the case of a lattice of oscillators with frequency dispersion given by a random distribution: in this case the source is a region that, due to random fluctuation, happens to have a higher average frequency than the surrounding oscillators. Again the improvement in frequency precision was reduced from $N^{-1}$ and instead scaled as $N_{C}^{-1}$ with $N_{C}$ the number of oscillators forming the core of the source.

Here we consider oscillators on a two-dimensional lattice. A two-dimensional lattice admits spirals: states with a $2\pi n$ phase winding. For such a phase winding the sum of phase differences between neighbors around any closed curve enclosing the spiral center is $2\pi n$ where $n$, the ``winding number'', counts the number of arms to the spiral. Since $n$ is constrained to take integer values and it is an observable property, it may only change if a spiral center migrates off the lattice or if two spirals of opposite winding number collide. Therefore $n$ is a conserved topological quantity which cannot be changed by normal continuous evolution for a single sprial. Spiral patterns are common features of synchronized oscillators on two-dimensional lattices \citep{Sakaguchi88,Blasius05}. In these states the spiral cores act as the sources of the phase waves. In this paper we will investigate analytically and numerically the frequency precision of a single spiral state. The state may form even in the case of identical oscillators, and we take advantage of this to focus mainly on this simpler case. We do not expect small frequency dispersion of the oscillators to significantly change the spiral state, except perhaps by favoring a particular location of the spiral core, or to change the frequency precision. We have confirmed this by numerical simulation.

It is convenient to use the abstract language of limit-cycle oscillators to discuss frequency sources. A limit-cycle oscillator, in the phase reduction description, is a device with state described by a periodic phase $\theta(t)$ (taking real values mod $2\pi$) advancing at a rate $\dot{\theta}(t)$ in time. When completely isolated, the oscillator rotates at a given \emph{natural frequency} $\dot{\theta}(t) = \omega$.
Every time a stochastic influence moves the state away from the limit cycle, until dissipation in the system causes it to relax back to the limit cycle, leading to a change in $\dot{\theta}$ from the uniform rotation. Thus noise is described by a stochastic function $\xi(t)$ acting on $\dot{\theta}$, and it is this noise that leads to the degradation of the frequency precision. The strength of this noise is denoted $f$.
The physical signal will be a periodic function of the phase $\theta$. The frequency precision is conveniently quantified by the inverse of the the width of the power spectrum of this signal in frequency space. 
In the synchronized state, we can describe the entire lattice by a single collective oscillator $\Theta(t)$ with mean frequency $\Omega$ and fluctuating component $\Xi(t)$ with strength $F$. The width of the signal obtained from ${\Theta}$ is the quantity that we wish to calculate. More specifically, we will calculate the ``relative frequency precision''  $R$ of the collective phase of a singe spiral synchronized state as compared to a single constituent oscillator. $R$ is defined as the ratio of the line widths of these two cases. 

The relationship of the strength of the noise $F$ acting on the collective phase to measures of the frequency precision of the system, such as the spectral density of fluctuations of a periodic function of the collective phase, is discussed in Ref.\ \cite{Cross12}. As discussed there, a white noise spectrum leads to \emph{diffusion} of the collective phase, and a broadening of the spectral peaks at the frequency of the oscillator and its harmonics, which are $\delta$-functions in the absence of noise, into Lorentzians with tails decreasing as $\delta\omega^{-2}$ with $\delta\omega$ the frequency offset from the peak frequency. Colored noise spectra will lead to different power laws for the decrease that are easily calculated from the stochastic equation for the collective phase. The broadening of the spectra of periodic signals measured from individual oscillators locked to the collective phase, or collections of such oscillators, will be similar. These oscillators will show additional fluctuations, which could be calculated from the exponential decay back onto the limit cycle given by the other modes of the system. However these fluctuations do not give a broadening of the $\delta$-function spectral peak of the ideal oscillator, since they do not affect the precision at which the frequency can be identified over the very long time assumed for the calculation of the spectrum, but rather contribute an additional background to the spectrum \cite{DemirMehrotra00}. This is analogous to thermal fluctuations in a crystal giving a Debye-Waller factor lowering the intensity of Bragg peaks in the diffraction pattern, but not broadening the peaks, which remain $\delta$-functions for an infinite crystal.

This paper is organized as follows: the model is defined in section~\ref{sec:model}, our analytic approach and approximations are described in section~\ref{sec:analytics}, results of numerical simulation are given in section~\ref{sec:numerics}, and in section~\ref{sec:discussion} these are discussed and conclusions drawn.

\section{Model} \label{sec:model}

The model used here is the same as that of Ref.~\citep{Cross12}. Each oscillator has an index and we define $\mathcal{N}_i$ as the set of indices of nearest-neighbor oscillators to $i$. Coupling is via a function $\Gamma(\chi)$. The equations of motion are therefore
\begin{eqnarray} \label{eqn:model}
\dot{\theta}_i = \omega_i + \xi_i(t) + \sum_{j \in \mathcal{N}_i}\Gamma(\theta_j - \theta_i).
\end{eqnarray}
The stochastic noise $\xi_i$ is assumed to be uncorrelated between different oscillators, with a white profile and strength $f$ so that $\left<\xi_i(t)\xi_j(t^\prime)\right> = f\delta(t - t^\prime)\delta_{ij}$. We offset the distribution of natural frequencies $\{\omega_i\}$ to have a mean of zero and define its width to be $\sigma$. For most of the paper we take $\sigma=0$. Our coupling function is one commonly used \citep{Paullet94,Cross12,Blasius05,Sakaguchi88,Sakaguchi08} and has degree of reactive coupling $\gamma$. We choose units of time so that it takes the form
\begin{eqnarray} \label{eqn:coupling}
\Gamma(\chi) = \sin\chi + \gamma\left(1 - \cos\chi\right).
\end{eqnarray}
Note that for $\gamma=0$ the coupling can be considered purely dissipative, since the sign of the coupling would reverse under time reversal $t\to -t,\theta_{i}\to-\theta_{i}$. This is the model studied by Kuramoto \cite{Kuramotobook}. On the other hand, the coupling term proportional to $\gamma$ does not change sign under time reversal, and so can be considered reactive (non-dissipative). This is discussed more fully by Topaj \emph{et al.}\ \cite{Topaj02}. The full coupling function contains both types of terms.

Reference \citep{Cross12} shows that the relative frequency precision is determined by the ratio of the collective and single phase noise strengths $R=F/f$. This is turn can be calculated \citep{Cross09,Cross12} by performing a linear perturbation analysis on Eq.~(\ref{eqn:model}), with no disorder, to give a linear Jacobian operator $\mathsf{J}$. $\mathsf{J}$ acts on vectors with components for each oscillator on the lattice. The result is
\begin{eqnarray} \label{eqn:CrossFoverf}
R = \frac{F}{f} = \frac{\mathbf{a}\cdot\mathbf{a}}{\left(\mathbf{a}\cdot\mathbf{s}\right)^2},\quad\mathbf{s}=(1,1,...,1),
\end{eqnarray}
with $\mathbf{a}$ the zero-eigenvalue adjoint eigenvector, defined by $\mathsf{J}^\dagger\cdot\mathbf{a}=0$. Note also that the normalization of $\mathbf{a}$ in Eq.~(\ref{eqn:CrossFoverf}) is unimportant. Here $\mathbf s$ is actually the zero-eigenvalue eigenvector of the Jacobian itself $\mathsf{J}\cdot\mathbf{s}=0$, normalized in a particular way (see Ref.~\cite{Cross12} for more details).

In order to perform the linear stability analysis we require a steady, no-disorder (i.e. $\xi_i=\omega_i=0\;\forall i$), solution to Eq.~(\ref{eqn:model}). With this we may construct $\mathsf{J}$ and solve for $\mathbf{a}$, finally calculating $R$. We follow this program analytically and numerically, testing the approximations used in the analytic approach against numerical results.

\section{Analytic approach} \label{sec:analytics}

\subsection{Spiral solution} \label{sec:spiralsolution}\label{sec:continuummodel}

We find an approximate form of the periodic spiral solution $\theta_{i}(t)=\theta_{i}+\Omega t$, with $\theta_{i}$ independent of $t$ and $\Omega$ the frequency, by approximating the lattice as a continuum.
Taking phase differences between neighboring oscillators tending to zero,
Eqs.~(\ref{eqn:model})~and~(\ref{eqn:coupling}) can be expanded to give this approximation. In the case of zero disorder we find
\begin{eqnarray} \label{eqn:continuummodel}
\dot{\theta}(\mathbf{r},t) = \nabla^2\theta + \gamma\left|\nabla\theta\right|^2.
\end{eqnarray}
$\theta$ is now a field over the lattice and $\mathbf{r}=(\rho,\varphi)$ is the two-dimensional position vector in polar coordinates. We will quantify this approximation from the form of $\theta(\mathbf{r},t)$ in a spiral solution below. Units of length have been set by taking lattice constant to be unity.

Following Ref.~\citep{Blasius05}, we look for perfectly-synchronized solutions ($\dot{\theta}=\Omega$) to Eq.~(\ref{eqn:continuummodel}) by linearizing using a Cole-Hopf transformation $\theta=\Omega t + \frac{1}{\gamma}\ln{G(\rho,\varphi)}$ giving
\begin{eqnarray} \label{eqn:G}
\nabla^2G = k^2G,\quad k = \sqrt{\Omega\gamma}.
\end{eqnarray}
For a spiral centered on $\rho=0$ we separate variables to solve Eq.~(\ref{eqn:G}), and require single-valuedness (mod $2\pi$) on the lattice $\varphi\rightarrow\varphi+2\pi$. We also require that all phases be real numbers. The solutions of Eq.~(\ref{eqn:G}) that satisfy these conditions are quantised with $n$ and are found to give
\begin{multline} \label{eqn:solution}
\theta(\rho,\varphi) = \Omega t + n\varphi + \frac{1}{\gamma}\ln K_{i n\gamma}(k\rho) + \text{const}.,\ n\in\mathbb{Z},
\end{multline}
where $K_\alpha(z)$ is the modified Bessel function of the second kind. The value of $k$ must be real and positive; the former condition requires $\Omega\gamma>0$, and so the frequency of the spiral is greater (resp.\ less) than the frequencies of the individual oscillators for $\gamma>0$ (resp. $\gamma<0$). $|n|$ represents the number of spiral arms, and we therefore do not form more general solutions of Eq.~(\ref{eqn:continuummodel}) by superposition of solutions with different $n$. In Ref.~\citep{Hagan82} it is shown that spirals obeying Eq.~(\ref{eqn:continuummodel}) are not stable unless $n=1$.

For large $k\rho$, Eq.~(\ref{eqn:solution})  can be approximated as 
\begin{align}\label{eqn:solutionapprox}
\theta\approx\Omega t + n\varphi - \frac{k\rho}{\gamma} - \frac{1}{2\gamma}\ln(2k\rho/\pi) + \mathcal{O}((k\rho)^{-1}),
\end{align}
demonstrating the spiral nature of the waves. The large $\rho$ asymptotic wave number is $k/\gamma$.
Since $\Omega\gamma>0,k>0$ the spiral waves are always \emph{outgoing} from the center. For definiteness and without loss of generality, we take $\gamma>0$ and so $\Omega>0$. In a finite system, there will be corrections to Eq.~(\ref{eqn:solutionapprox}) approaching the boundaries: we expect these to be significant within a distance $\sim k^{-1}$ of the boundaries.

The continuum approximation is good for small phase differences between adjacent oscillators, which requires large $\gamma\rho$ and small $k/\gamma$. Note that since $\omega_i = 0,\ \forall i$, the only feature driving oscillations is the topological phase winding. One therefore expects the oscillation frequency $\Omega$ to be set in the central region of the spiral, not accessible to the continuum approximation.

\subsection{Jacobian and adjoint eigenvector} \label{sec:eigenvector}

We follow the method used in Ref.~\citep{Cross12} to calculate the zero-eigenvalue adjoint eigenvector $\mathbf a$ of $\mathsf{J}^{\dagger}$. The Jacobian matrix $\mathsf{J}$ describes the linear phase dynamics about the fixed point phases $\theta_{i}$ which define the limit cycle. The components of $\mathsf{J}$ are
\begin{eqnarray} \label{eqn:forwardJacobian}
J_{ij}= \begin{cases} \Gamma'(\theta_{j}-\theta_{i}) & \mbox{if } j\in\mathcal{N}_i \\
-\sum_{l\in\mathcal{N}_i}\Gamma'(\theta_{l}-\theta_{i}) & \mbox{if } j = i \end{cases}
\end{eqnarray}
with all other components zero. Our analysis will require the adjoint $J^{\dagger}_{\ ij}=J_{ji}$.

The continuum approximation used to find the solution Eq.~(\ref{eqn:solution}) requires small phase differences between neighboring oscillators. When this is valid the interaction function Eq.~(\ref{eqn:coupling}) can be approximated \citep{Blasius05}
\begin{eqnarray} \label{eqn:couplingapprox}
\Gamma(\chi) \approx \gamma^{-1}(e^{\gamma\chi}-1)\Rightarrow\Gamma'(\chi) \approx e^{\gamma\chi},
\end{eqnarray}
as was also used in Ref.~\citep{Cross12}. This approximation agrees with the full expression for $\Gamma(\chi)$ up to second order in small $\chi$. It can be used \citep{Blasius05} to implement a Cole-Hopf transformation on the lattice, which then leads to Eq.~(\ref{eqn:G}) taking the continuum limit. With this approximate interaction function, the adjoint Jacobian takes the form
\begin{eqnarray} \label{eqn:discreteJacobian}
J^\dagger_{ij} \approx \begin{cases} e^{\gamma(\theta_{i} - \theta_j)} & \mbox{if } j\in\mathcal{N}_i \\
-\sum_{l\in\mathcal{N}_i}e^{\gamma(\theta_l-\theta_i)} & \mbox{if } j = i. \end{cases}
\end{eqnarray}

In order to follow Ref.~\cite{Cross12} in solving for $\mathbf{a}$ we would need to make a Cole-Hopf transformation $Q_i = \exp(\gamma\theta_i)$ in Eq.~(\ref{eqn:discreteJacobian}). However note that while Eq.~(\ref{eqn:solution}) is single-valued (mod $2\pi$) as $\varphi\rightarrow\varphi + 2\pi$, such a transformation would produce a multi-valued $Q_i$. We will see that this leads to an unacceptable solution for $\mathbf{a}$. One could try to make $Q_i$ single-valued by restricting the range of allowed values for $\varphi$ to $[0,2\pi)$ before making the Cole-Hopf transformation. But such a restiction would not be compatible with the approximation of small phase differences used in Eq.~(\ref{eqn:couplingapprox}) as it would lead to $\mathcal{O}(2\pi n)$ phase differences across the $\varphi=0$ line. The linear $\varphi$-dependence of Eq.~(\ref{eqn:solution}) therefore prevents us from proceeding in this way.

In order to make progress, we define $\psi(\rho)$ as the radially-dependent part of Eq.~(\ref{eqn:solution}) so that
\begin{equation}
\theta(\rho,\varphi) = \Omega t + n\varphi + \psi(\rho).\label{eqn:introducepsi}
\end{equation}
We then note that a pair of nearest-neighbor oscillators $(i,j)$ at large-$\rho$ (where Eq.~(\ref{eqn:solution}) is a good approximation) satisfy $\varphi_j = \varphi_i + \mathcal{O}(\rho_j^{-1})$. As such their phase differences are given by

\begin{eqnarray}\label{eqn:phasediffs}
\theta_j - \theta_i = \psi_j - \psi_i + \mathcal{O}(\rho_j^{-1}).
\end{eqnarray}
These expressions for phase differences remain small for all $\varphi$. Combining these approximations, Eqs.~(\ref{eqn:couplingapprox},\ref{eqn:phasediffs}), the adjoint Jacobian becomes
\begin{eqnarray} \label{eqn:approxdiscreteJacobian}
J^\dagger_{ij} \approx \begin{cases} e^{\gamma(\psi_i - \psi_j)} & \mbox{if } j\in\mathcal{N}_i \\
                      -\sum_{l\in\mathcal{N}_i}e^{\gamma(\psi_l-\psi_i)} & \mbox{if } i = j. \end{cases}
\end{eqnarray}

Now Eq.~(\ref{eqn:approxdiscreteJacobian}) is the same functional form as found in Ref.~\citep{Cross12}, with no $\varphi$-dependence. As such, one can safely perform a Cole-Hopf transformation $q_i=\exp(\gamma\psi_i)$, which is single valued. Under this transformation the Jacobian takes the form
\begin{eqnarray}
J^\dagger_{ij} \approx \begin{cases} q_i/q_j & \mbox{if } j\in\mathcal{N}_i \\
                      -\sum_{l\in\mathcal{N}_i}q_l/q_i & \mbox{if } i = j \end{cases}
\end{eqnarray}
from which $a_i \approx q_i^2$ is found by inspection. Here we see why $Q_i$ being multi-valued is unacceptable, as it would have produced a multi-valued adjoint eigenvector. Inverting the Cole-Hopf tranformation and inserting the solution (\ref{eqn:solution}) we find the adjoint eigenvector $a$ to  be given by
\begin{eqnarray} \label{eqn:eigenvector}
a \propto \left[K_{in\gamma}(kr)\right]^{2}.
\end{eqnarray}

The $\varphi$ independence of the adjoint eigenvector Eq.~(\ref{eqn:eigenvector}) is expected. Note that the zero-eigenvalue eigenvector to the Jacobian itself (the continuum version of the vector $\mathbf {s}$ in Eq.~(\ref{eqn:CrossFoverf})) is a constant over the lattice, corresponding to the zero mode of a small rotation of all the phases together, and so has no $\varphi$ dependence in an infinite system. Since $\mathbf{a}^{\dagger}\cdot\mathbf{a}\ne0$, the zero-eigenvalue adjoint eigenvector must have the same azimuthal symmetry. In a finite system there will be corrections approaching the boundaries.

We can verify the consistency of this result with a direct continuum calculation.
Within the continuum approximation, the Jacobian operator from Eq.~(\ref{eqn:continuummodel}) is
\begin{equation}
J=\nabla^{2} + 2\gamma(\vec\nabla\theta)\cdot\vec\nabla,
\end{equation}
with $\theta$ the steady state spiral solution, Eq.~(\ref{eqn:solution}),
giving the adjoint
\begin{eqnarray} \label{eqn:continuumadjoint}
\hat{J}^\dagger = \nabla^2 - 2\gamma\vec\nabla\theta\cdot\vec\nabla - 2\gamma\nabla^2\theta.
\end{eqnarray}
The continuum approximation to $\mathbf a$ is given by the zero-eigenvalue eigenvector
\begin{equation} \label{eqn:continuumadjoint=0}
\hat{J}^\dagger a=0.
\end{equation}
Similar equations have been studied by T\"onjes and Blasius \cite{Tonjes09}.
With $\theta$ of the form Eq.~(\ref{eqn:introducepsi}), it can be verified by direct substitution that  $a(\rho)=e^{2\gamma\psi(\rho)}$ satisfies Eq.~(\ref{eqn:continuumadjoint=0}). Using the explicit form of $\psi(\rho)$ from Eq.~(\ref{eqn:solution}) again gives Eq.~(\ref{eqn:eigenvector}).

\subsection{Relative frequency precision} \label{sec:Foverf}

\begin{figure}[tbh]
\begin{center}
\includegraphics[width=0.9\columnwidth]{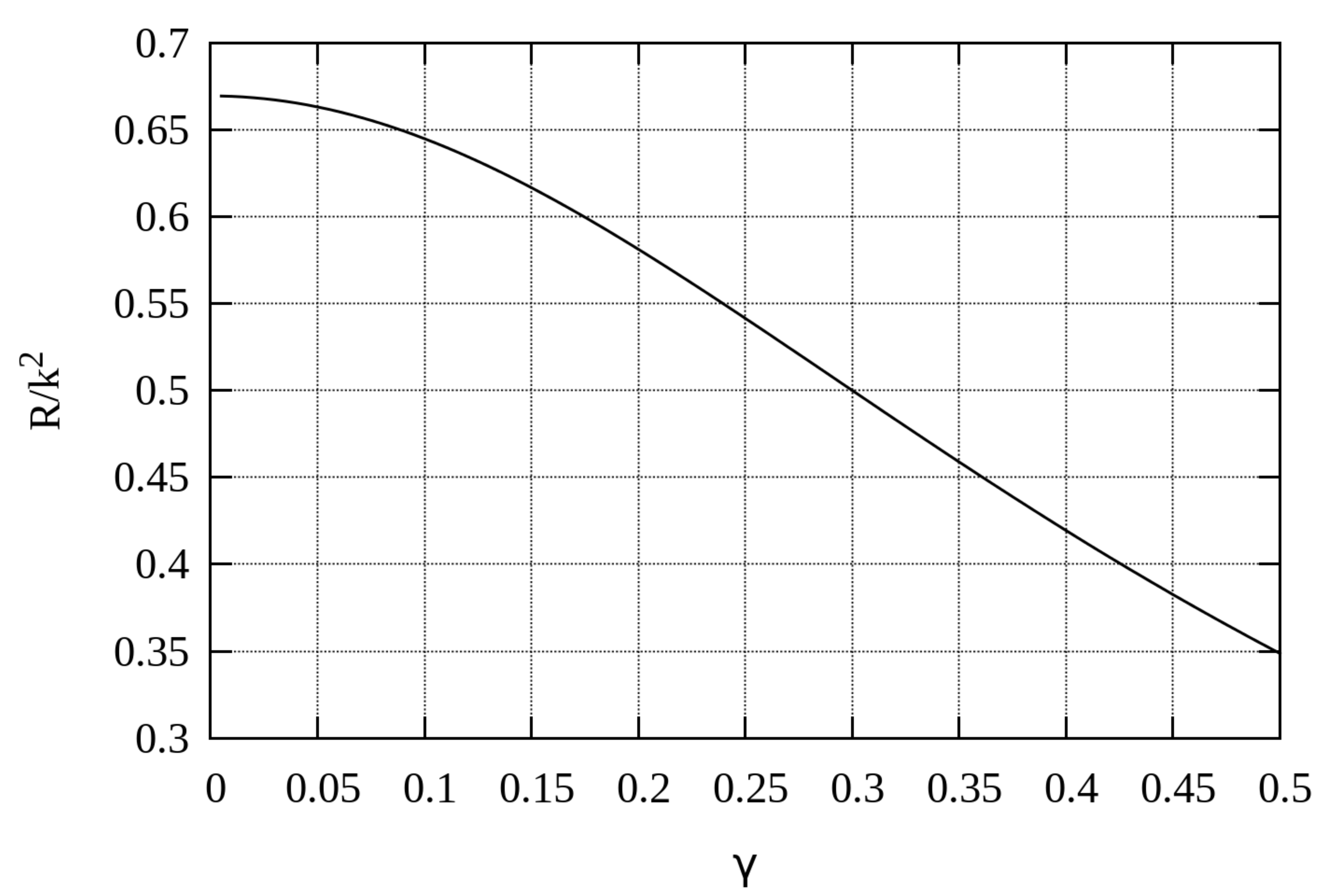}
\caption{\label{fig:Roverk2} Plot of $R/k^{2}$ derived from the analytic continuum approximation, with $R$ the relative frequency precision and $k=\sqrt{\Omega\gamma}$ with $\Omega$ the spiral frequency, as a function of the reactive coupling constant $\gamma$.}
\end{center}
\end{figure}

In the continuum approximation the relative frequency precision Eq.~(\ref{eqn:CrossFoverf}) approximates to
\begin{equation} \label{eqn:continuumFoverf}
R \approx \frac{\int a^2\mathrm{d}\mathbf{r}} {\left(\int a\mathrm{d}\mathbf{r}\right)^2}.
\end{equation}
Using Eq.~(\ref{eqn:eigenvector}) and integrating over an infinite domain gives
\begin{equation} \label{eqn:Rvsgamma}
R=k^{2}F(\gamma),
\end{equation}
where $F(\gamma)$ can be expressed in terms of a Meijer's G-function. The function is plotted in Fig.~\ref {fig:Roverk2}. The value for $\gamma\to 0$ is $F(0)=7\zeta(3)/4\pi\simeq0.6696$.
The continuum approximation is good for small $\gamma$. In a finite system of $N$ oscillators, since the adjoint eigenvector decays as $e^{-2k\rho}/\rho$ for large $\rho$, there will also be corrections from the boundaries when $k\sqrt N$ is insufficiently large.

\subsection{Spiral frequency}

The analytic prediction for the relative frequency precision of a spiral depends on the, as yet unspecified, dependence of $k$ (equivalently, $\Omega$) on $\gamma$. In a pioneering paper Hagan \citep{Hagan82} investigated spiral formation in continuum amplitude-phase equations for collective oscillations. Away from the core of the spiral a nonlinear phase description applies, which is isomorphic to our continuum phase model (\ref{eqn:continuummodel}). In the core of the spiral the amplitude becomes suppressed, going to zero at the center of the spiral. As in our work, the frequency of the spiral is arbitrary in the large-distance phase description: Hagan finds  the frequency is set by matching to the inner, core region \footnote{Hagan's method of solving the nonlinear phase equation is different to our: he defined ``middle'' and ``outer'' regions with different solution forms in the two regions, with some constants set by matching between these two regions. Both regions are described by our Bessel function solution derived from the Cole-Hopf transformation.}. By matching between the phase-only region, and the inner core region Hagan was able to find the relationship
\begin{eqnarray} \label{eqn:hagank}
k(\gamma) = 2\exp\left(-\frac{\pi}{2\gamma} + \alpha\right),
\end{eqnarray}
where the constant $\alpha$ is determined by the core structure. We will use this expression, but expect a different value of $\alpha$ due to the different, discrete, nature of our cores. The parameter $\alpha$ will be determined by fitting $\Omega=k^{2}/\gamma$ to the numerical results.

\section{Numerical simulations} \label{sec:numerics}

\subsection{Approach}

\begin{figure}[tbh]
\begin{center}
\includegraphics[width=0.9\columnwidth]{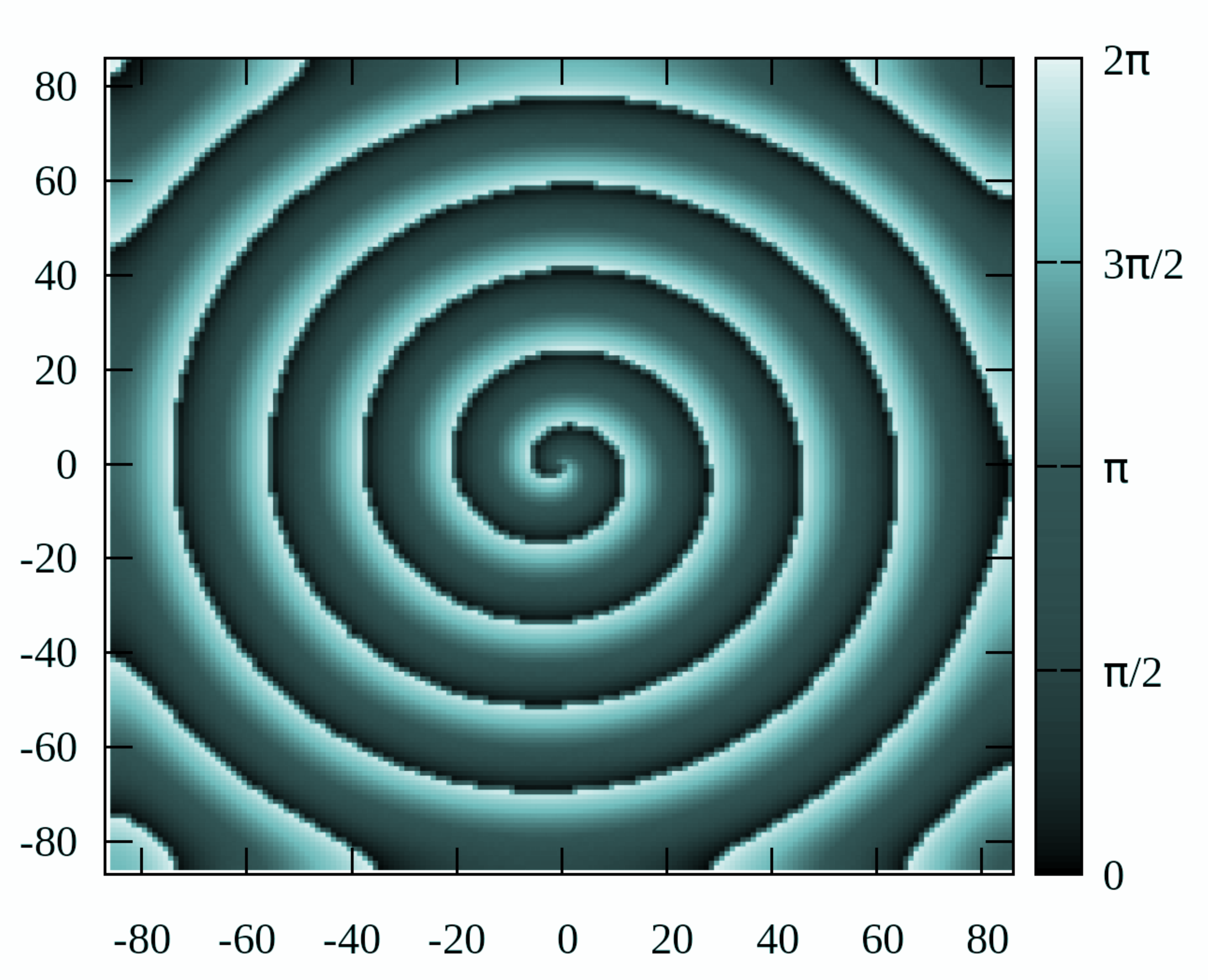}
\caption {\label{fig:spiral} Spiral on lattice of $N=29,929$ and $\gamma=0.45$ after evolution to convergence of $r<10^{-3}$. Colored squares are individual oscillators with phase given by the colorbar.
}
\end{center}
\end{figure}

In order to numerically investigate the relative frequency precision $R$ and frequency $\Omega$ of the synchronized spiral state we simulate square lattices of $N$ oscillators with square unit cells.  We will assume, for simplicity, that the spiral is centered on the lattice.
For an initial condition we use Eq.~(\ref{eqn:solution}), with an estimated value of $k$.
We evolve the system using numerical integration until convergence; Fig.~\ref{fig:spiral} shows the spiral resulting from one simulation. From a converged state ($\omega_i\approx\Omega\;\forall i$) $\Omega$ can be found by summing the equations of motion (\ref{eqn:model}). Noting that there are four nearest neighbors per oscillator this gives 
\begin{equation}
\label{eqn:Omega}
\Omega=4\gamma\left<1-\cos(\theta_j - \theta_i)\right>_{j\in\mathcal{N}_i,\;\forall i}.
\end{equation}

To assess convergence we write, when close to synchrony, $\dot{\theta}_i = \Omega(1 + r\delta_i)$ for some set $\{\delta_i\}$ with zero mean and unit standard deviation. As such, $r$ is the ratio of the standard deviation to the mean of the $\{\dot{\theta}_i\}$, an easily calculable quantity that quantifies the distance from convergence. We define convergence for our simulations when $r < 10^{-3}$, all simulations were run until this condition was met. We investigated a parameter space $14884 \le N \le 29929$, $n=1$, and $0.1 \le\gamma\le 0.45$.
Smaller values of $\gamma$ produce slower-moving waves and therefore slower convergence. Larger values of $\gamma$ cause spirals to migrate around the lattice and never settle \citep{Hagan82}.

When a converged solution has been reached the discrete adjoint Jacobian is constructed, $\mathsf{J}^\dagger\cdot\mathbf{a}=0$ is numerically inverted to find $\mathbf{a}$, and $R$ is calculated.

\subsection{Results}

\begin{figure}[tbh]
\begin{center}
\includegraphics[width=0.9\columnwidth]{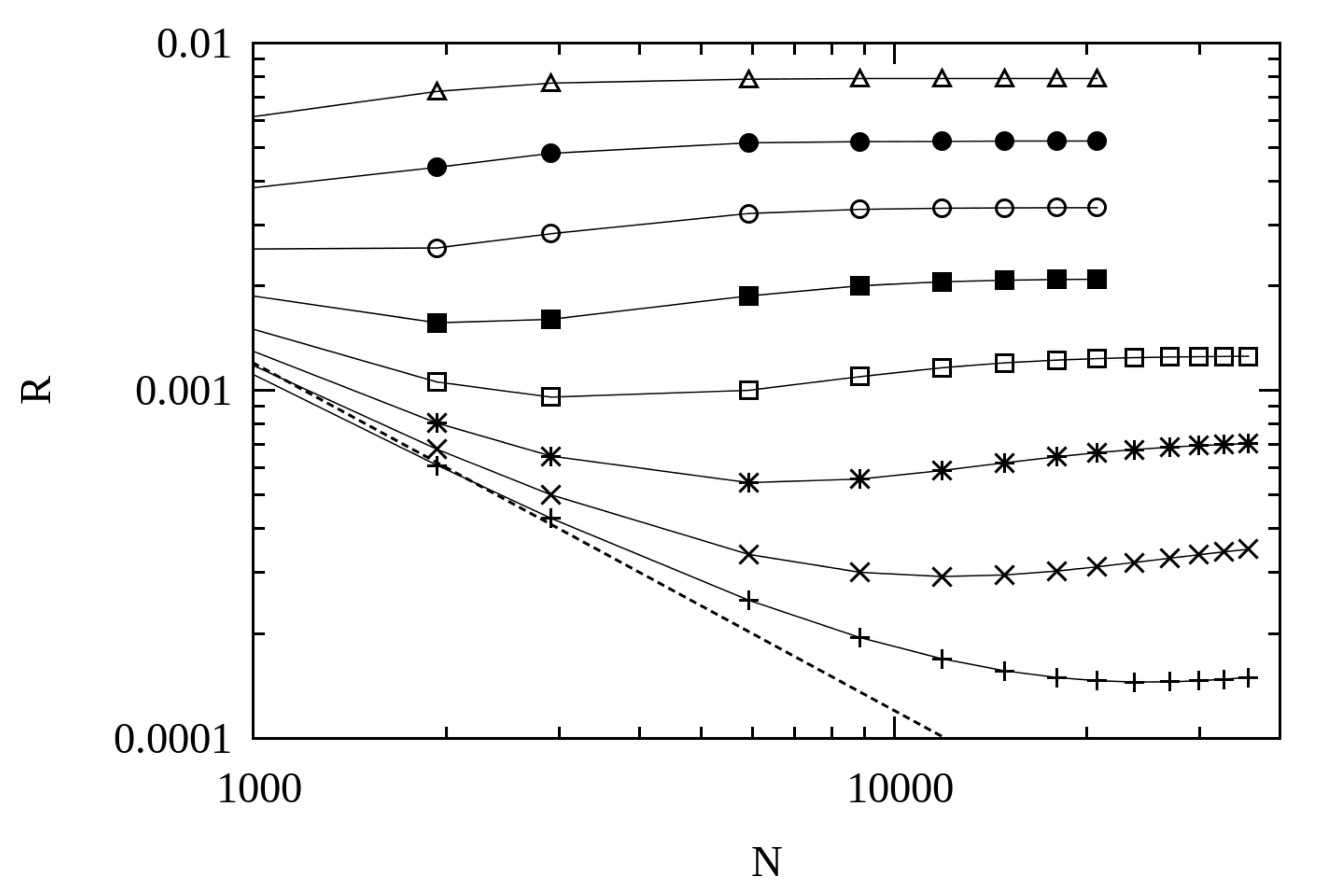}
\caption {\label{fig:RvsN} Relative frequency precision $R$ as function of the number of oscillators $N$ in a square lattice for a selection of values of the reactive coupling constant $\gamma$ evenly spaced between 0.29 and 0.43 ($\gamma$ increases up the diagram). For small $\gamma$ and $N$ the data are consistent with $R\propto N^{-1}$ (dashed line), as would be found for the model without reactive coupling. For large enough $N$, the data is consistent with a value of $R$ independent of system size.}
\end{center}
\end{figure}

A key result of our analysis is that for large enough system sizes $N$ and $\gamma\ne 0$, the improvement in frequency precision becomes independent of the system size, corresponding to the localized nature of the source of the waves, as was found in Refs.~\cite{Masuda10,Cross12} for nontopological sources due to inhomogeneities in the lattice. This is shown in Fig.~\ref{fig:RvsN}. The size needed to reach the large system limit depends strongly on $\gamma$, increasing as $\gamma$ decreases: this is consistent with the increase in the decay length $(2k)^{-1}$ of the adjoint eigenvector and the rapid decrease of $k$ with $\gamma$ predicted by Eq.~(\ref{eqn:hagank}).

Fig.~\ref{fig:RvsN} shows only a subset of the numerical data to demonstrate the variation with $N$. For smaller $\gamma$ the system sizes accessible to us are insufficient to reach this asymptotic large $N$ limit. In order to compare with analytic expressions we therefore define a large-$N$ subset of the data for which the edge effects are ignorable. The ``large-$N$'' data are defined as those simulations satisfying $k\sqrt{N}>4$.

\begin{figure}[tbh]
\begin{center}
\includegraphics[width=0.9\columnwidth]{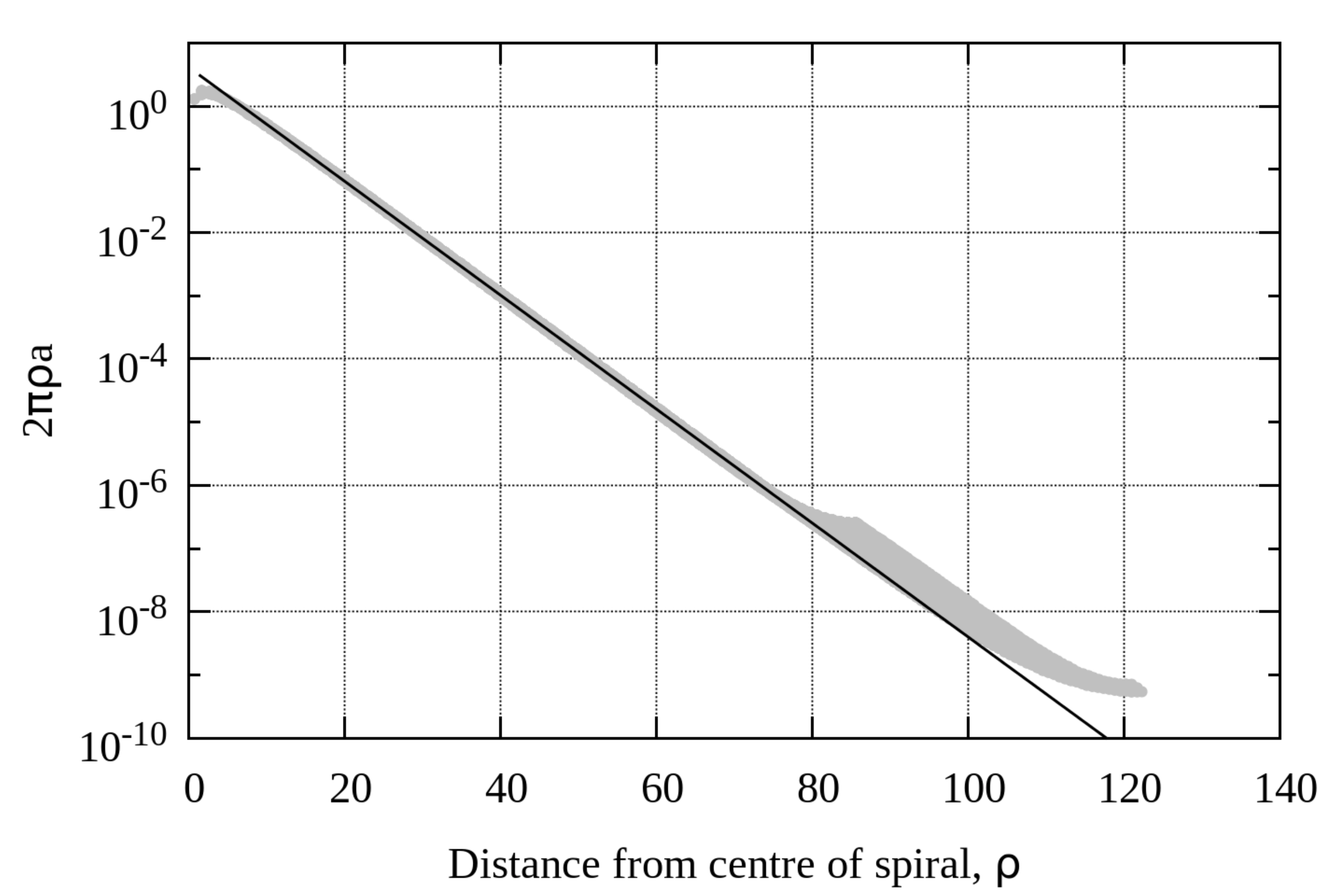}
\caption {\label{fig:eigenvector} Adjoint eigenvector on a lattice of $N=29,929$ oscillators and $\gamma=0.4$: $2\pi \rho a(\rho)$ is plotted on a logarithmic scale as a function of the distance $\rho$ from the center of the spiral, which is chosen as the interpolated point with maximum $a$. Numerical data are shown in grey; black curve is Eq.~(\ref{eqn:eigenvector}) fitted for $k$ and a multiplicative constant. Fits performed against points $\rho<80$ to avoid the edge effects which are apparent for $\rho\gtrsim80$.}
\end{center}
\end{figure}

To compare with the analytic predictions we first show in Fig.~\ref{fig:eigenvector} the eigenvector $a(\rho)$ for a particular simulation run. The spiral center of the converged solution has been defined as the point of maximum $a$ and the analytic expression (\ref{eqn:eigenvector}) has been fitted to the data for $k$ and a multiplicative constant. The good correspondence and $\varphi$-independence justifies the expression for $a$ (\ref{eqn:eigenvector}). The fit is good for $\rho\gtrsim 2$ (two lattice spacings) lending credence to the continuum approach. Additionally, values of $k$ fitted in this way agree well with the values of $k=\sqrt{\Omega\gamma}$ numerically calculated from Eq.~(\ref{eqn:Omega}).

\begin{figure}[tbh]
\begin{center}
\includegraphics[width=0.9\columnwidth]{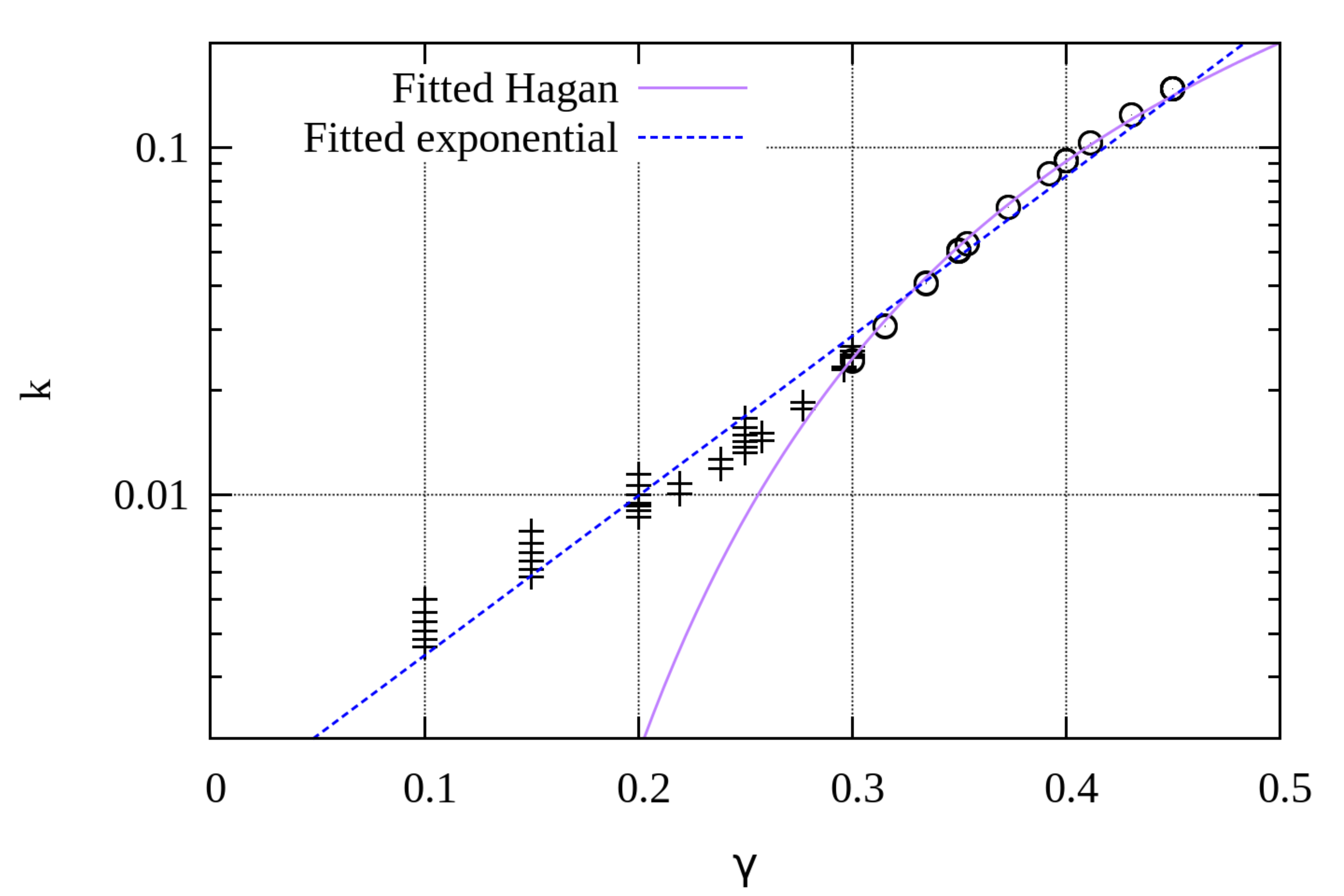}
\caption {\label{fig:kvsgamma} Spiral parameter $k$ as a function of reactive coupling constant $\gamma$ for system sizes $14884 \le N \le 29929$: circles -- large-$N$ numerical data for which edge effects are ignorable ($k\sqrt{N}>4$); pluses -- remaining numerical data; solid line --Hagan relationship (\ref{eqn:hagank}) fitted to circles giving $\alpha\approx0.840$; blue line -- empirical exponential fitted to all data points. }
\end{center}
\end{figure}

Figure \ref{fig:kvsgamma} plots these numerical values of $k$ as a function of $\gamma$. The ``large-$N$'' data (with larger $\gamma$, denoted by circles) are those for which edge effects are small. These are well fit by the Hagan expression (\ref{eqn:hagank}) with the matching constant $\alpha\approx0.840$. The complete data set is actually quite well fit by a simple exponential relationship $k=pe^{q\gamma}$ with $p\approx1.20\e{-3}$ and $q\approx10.6$. However, it is clear that when $\gamma=0$ we must have $\Omega=k=0$ and therefore the true relationship cannot be an exponential, although this fit may be useful to give approximate values of $k(\gamma)$ in finite systems.

\begin{figure}[tbh]
\begin{center}
\includegraphics[width=0.9\columnwidth]{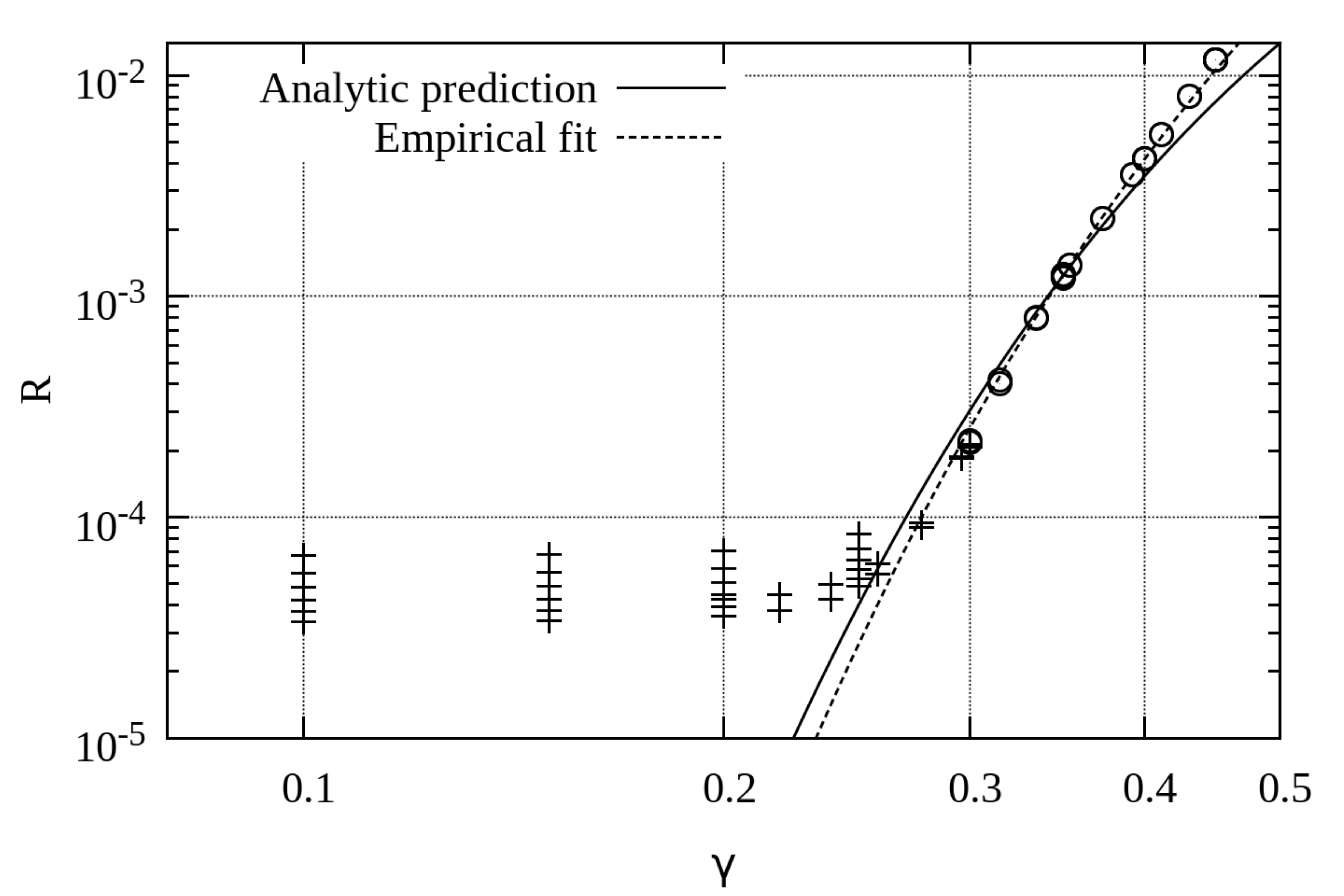}
\caption {\label{fig:Rvsgamma} Relative frequency precision $R$ as a function of $\gamma$ for system sizes $14884 \le N \le 29929$: circles -- large-$N$ numerical data for which edge effects are ignorable ($k\sqrt{N}>4$); pulses -- remaining numerical data; solid line -- analytic prediction Eqs.~(\ref{eqn:Rvsgamma},\ref{eqn:hagank}) with $\alpha\approx0.840$; dashed line -- empirical fit for $R=ak^b$ with Eq.~(\ref{eqn:hagank}) for $k(\gamma)$.}
\end{center}
\end{figure}

With $k(\gamma)$ given by Eq.~(\ref{eqn:hagank}) and the fitted value of $\alpha=0.84$ we compare in Fig.~\ref{fig:Rvsgamma} the predictions of the analytic theory of Section \ref{sec:analytics} with the values of $R$ as a function of gamma given by Eq.~(\ref{eqn:CrossFoverf}) and the adjoint eigenvector from the simulations. Deviations of the data from the analytic predictions are expected at smaller $\gamma$ due to finite size effects (the noise reduction factor becomes $N^{-1}$ independent of $\gamma$) and at larger values of $\gamma$ due to discrete lattice effects. There is good agreement over the intermediate regime of $\gamma$, spanning two orders of magnitude in the noise suppression factor. Note that the analytic prediction contains no additional fit parameters once the constant $\alpha$ is determined from spiral frequency $\Omega(\gamma)$ in Fig.~\ref{fig:kvsgamma}.  We also show for camparison an empirical power law fit  $R\approx ak^b$ using the same expression for $k(\gamma)$: fitting such a power law to the large-$N$ data produces $a\approx0.70$ and $b\approx2.14$.

A single simulation was run with finite disorder in natural frequencies. Using a uniform distribution of $\{\omega_i\}$ centered on zero with $\sigma=0.2$ it was found that convergence to a steady spiral state satisfying $r<10^{-3}$ was reached in a comparable time to the case without disorder. This confirms that our results  apply to real-world cases with nonzero-$\sigma$.

\section{Discussion} \label{sec:discussion}

We have presented results for the enhancement in frequency precision of a two dimensional array of oscillators synchronized in a state of propagating spiral waves. Our results are summarized in Fig.\  \ref {fig:Rvsgamma}.
In the presence of propagation $\gamma\ne 0$ we find a frequency precision that is independent of the size of the system for large enough systems, unlike the case of no propagation $\gamma=0$ where the frequency precision improves in proportion to the total number of oscillators, as would be given by averaging independent noise fluctuations of the individual oscillators.
Our result is consistent with the intuition that a core region of oscillators determines the frequency of the spiral state, and the noise acts on this core region to degrade the frequency precision, so that noise averaging is less effective. A similar result was found in Refs.\ \cite{Masuda10,Cross12} for localized sources that arise due to inhomogeneity in the lattice of oscillators. In the present case the localized spiral source may form depending on initial conditions even in the absence of inhomogeneity, and is topologically robust so then persists indefinitely. This allows us to develop an analytic understanding of the numerical results in the simpler, regular system.
For small $\gamma$ the improvement in frequency precision can be significant, down to $R\approx10^{-5}$ in the lattice sizes we investigated. For larger $\gamma$ the spiral core shrinks, and the improvement in frequency precision is less effective, approaching $R\approx10^{-2}$ for $\gamma=0.5$.   For even larger values of $\gamma$ the spirals migrate through the lattice  \citep{Hagan82} (we observed this in several simulations when $\gamma\gtrsim0.6$), leading to a unsynchronized state.

We have used a phase reduction model to describe the oscillators. Since the key effect is the long time diffusion of the collective phase, our main conclusions will apply with small modifications to more general models such as amplitude-phase models, or full descriptions of the individual oscillator equations. The core region of the spirals, where the oscillator phases deviate from the continuum phase description, will be depend on the details of the model. For small $\gamma$, where the continuum analytic approximation applies, this will change the constant $\alpha$ in Eq.~(\ref{eqn:hagank}), which is given by matching to the core region, but not other results that depend only on the properties of the spiral far from the core.

In the presence of weak disorder, we expect our  conclusions will continue to hold, as we have verified by numerical simulation. With this disorder, however, the spiral core may migrate to a preferred location in the array driven by the disorder but impeded by lattice pinning. For stronger disorder we expect a competition between the localized sources of Ref.\ \cite{Cross12} and the spiral sources. Typically in such nonlinear wave systems it is the \emph{higher} frequency sources that win such a competition for $\gamma>0$ \cite{Cross93}. This can be understood from the motion of the shocks that form between the regions of waves emanating from the different sources in a Cole-Hopf analysis of the continuum description. Discrete lattice effects may tend to pin the shocks, leading to the possibility of more than one source persisting to long times, again giving an unsynchronized state.

An interesting extension of our work would be to the multi-spiral states that are often seen in numerical simulations from random initial conditions. Even in a system without disorder, the question arises of whether such a state is a periodic (synchronized) state of stationary spirals, or a chaotic state of dynamic spirals as is seen in other spiral wave systems for some parameter values \cite{Aranson02}. In the presence of disorder the individual spirals, even if stationary, might be expected to have different frequencies, determined by the local environment of their cores, giving an unsynchronized state. On the other hand, the interaction of the spirals through the shocks between them might tend to lock the frequencies, returning the system to a synchronized state.

A further extension would be to study larger noise strengths. As in Ref.\ \cite{Cross12}, our analysis of the effect of noise on the frequency precision is restricted to the small noise limit, where the noise acts on the free collective phase variable, but is ineffective in jumping the system over barriers such as would lead to noise driven diffusion of the spiral core. The rate of such processes scale as $e^{-\Delta/f}$ with $\Delta$ a measure of the barrier height, and $f$ the noise strength, and so becomes vanishingly small for small enough noise.

\begin{acknowledgments}
This material is based upon work supported by the National Science Foundation under Grant No.\ DMR-1003337. JMAA is supported by the Master and Fellows of Corpus Christi College and the California Institute of Technology.\end{acknowledgments}

\bibliography{references} 

\begin{thebibliography}{20}
\expandafter\ifx\csname natexlab\endcsname\relax\def\natexlab#1{#1}\fi
\expandafter\ifx\csname bibnamefont\endcsname\relax
  \def\bibnamefont#1{#1}\fi
\expandafter\ifx\csname bibfnamefont\endcsname\relax
  \def\bibfnamefont#1{#1}\fi
\expandafter\ifx\csname citenamefont\endcsname\relax
  \def\citenamefont#1{#1}\fi
\expandafter\ifx\csname url\endcsname\relax
  \def\url#1{\texttt{#1}}\fi
\expandafter\ifx\csname urlprefix\endcsname\relax\def\urlprefix{URL }\fi
\providecommand{\bibinfo}[2]{#2}
\providecommand{\eprint}[2][]{\url{#2}}

\bibitem[{\citenamefont{Winfree}(2001)}]{Winfree01}
\bibinfo{author}{\bibfnamefont{A.}~\bibnamefont{Winfree}},
  \emph{\bibinfo{title}{The Geometry of Biological Time}}, Interdisciplinary
  applied mathematics: Mathematical biology (\bibinfo{publisher}{Springer},
  \bibinfo{year}{2001}), ISBN \bibinfo{isbn}{9780387989921}.

\bibitem[{\citenamefont{Ekinci and Roukes}(2005)}]{Ekinci05}
\bibinfo{author}{\bibfnamefont{K.~L.} \bibnamefont{Ekinci}} \bibnamefont{and}
  \bibinfo{author}{\bibfnamefont{M.~L.} \bibnamefont{Roukes}},
  \bibinfo{journal}{Rev. Sci. Instrum.} \textbf{\bibinfo{volume}{76}},
  \bibinfo{pages}{061101} (\bibinfo{year}{2005}).

\bibitem[{\citenamefont{Kenig et~al.}(2012)\citenamefont{Kenig, Cross,
  Lifshitz, Karabalin, Villanueva, Matheny, and Roukes}}]{Kenig12}
\bibinfo{author}{\bibfnamefont{E.}~\bibnamefont{Kenig}},
  \bibinfo{author}{\bibfnamefont{M.~C.} \bibnamefont{Cross}},
  \bibinfo{author}{\bibfnamefont{R.}~\bibnamefont{Lifshitz}},
  \bibinfo{author}{\bibfnamefont{R.~B.} \bibnamefont{Karabalin}},
  \bibinfo{author}{\bibfnamefont{L.~G.} \bibnamefont{Villanueva}},
  \bibinfo{author}{\bibfnamefont{M.~H.} \bibnamefont{Matheny}},
  \bibnamefont{and} \bibinfo{author}{\bibfnamefont{M.~L.}
  \bibnamefont{Roukes}}, \bibinfo{journal}{Phys. Rev. Lett.}
  \textbf{\bibinfo{volume}{108}}, \bibinfo{pages}{264102}
  (\bibinfo{year}{2012}).

\bibitem[{\citenamefont{Pikovsky et~al.}(2001)\citenamefont{Pikovsky,
  Rosenblum, and Kurths}}]{Pikovsky01}
\bibinfo{author}{\bibfnamefont{A.}~\bibnamefont{Pikovsky}},
  \bibinfo{author}{\bibfnamefont{M.}~\bibnamefont{Rosenblum}},
  \bibnamefont{and} \bibinfo{author}{\bibfnamefont{J.}~\bibnamefont{Kurths}},
  \emph{\bibinfo{title}{Synchronization: A Universal Concept in Nonlinear
  Sciences}}, Cambridge Nonlinear Science Series (\bibinfo{publisher}{Cambridge
  University Press}, \bibinfo{year}{2001}).

\bibitem[{\citenamefont{Chang et~al.}(1997)\citenamefont{Chang, Cao, Mishra,
  and York}}]{Chang97}
\bibinfo{author}{\bibfnamefont{H.-C.} \bibnamefont{Chang}},
  \bibinfo{author}{\bibfnamefont{X.}~\bibnamefont{Cao}},
  \bibinfo{author}{\bibfnamefont{U.}~\bibnamefont{Mishra}}, \bibnamefont{and}
  \bibinfo{author}{\bibfnamefont{R.}~\bibnamefont{York}},
  \bibinfo{journal}{IEEE Trans. Microwave Theory Tech.}
  \textbf{\bibinfo{volume}{45}}, \bibinfo{pages}{604 } (\bibinfo{year}{1997}),
  ISSN \bibinfo{issn}{0018-9480}.

\bibitem[{\citenamefont{Needleman et~al.}(2001)\citenamefont{Needleman,
  Tiesinga, and Sejnowski}}]{Needleman01}
\bibinfo{author}{\bibfnamefont{D.~J.} \bibnamefont{Needleman}},
  \bibinfo{author}{\bibfnamefont{P.}~\bibnamefont{Tiesinga}}, \bibnamefont{and}
  \bibinfo{author}{\bibfnamefont{T.~J.} \bibnamefont{Sejnowski}},
  \bibinfo{journal}{Physica D} \textbf{\bibinfo{volume}{155}},
  \bibinfo{pages}{324} (\bibinfo{year}{2001}).

\bibitem[{\citenamefont{Topaj and Pikovsky}(2002)}]{Topaj02}
\bibinfo{author}{\bibfnamefont{D.}~\bibnamefont{Topaj}} \bibnamefont{and}
  \bibinfo{author}{\bibfnamefont{A.}~\bibnamefont{Pikovsky}},
  \bibinfo{journal}{Physica D} \textbf{\bibinfo{volume}{170}},
  \bibinfo{pages}{119} (\bibinfo{year}{2002}).

\bibitem[{\citenamefont{Sakaguchi et~al.}(1988)\citenamefont{Sakaguchi,
  Shinomoto, and Kuramoto}}]{Sakaguchi88}
\bibinfo{author}{\bibfnamefont{H.}~\bibnamefont{Sakaguchi}},
  \bibinfo{author}{\bibfnamefont{S.}~\bibnamefont{Shinomoto}},
  \bibnamefont{and} \bibinfo{author}{\bibfnamefont{Y.}~\bibnamefont{Kuramoto}},
  \bibinfo{journal}{Prog.Theor. Phys.} \textbf{\bibinfo{volume}{79}},
  \bibinfo{pages}{1069} (\bibinfo{year}{1988}).

\bibitem[{\citenamefont{Masuda et~al.}(2010)\citenamefont{Masuda, Kawamura, and
  Kori}}]{Masuda10}
\bibinfo{author}{\bibfnamefont{N.}~\bibnamefont{Masuda}},
  \bibinfo{author}{\bibfnamefont{Y.}~\bibnamefont{Kawamura}}, \bibnamefont{and}
  \bibinfo{author}{\bibfnamefont{H.}~\bibnamefont{Kori}}, \bibinfo{journal}{New
  J. Phys.} \textbf{\bibinfo{volume}{12}}, \bibinfo{pages}{093007}
  (\bibinfo{year}{2010}).

\bibitem[{\citenamefont{Cross and Greenside}(2009)}]{Cross09}
\bibinfo{author}{\bibfnamefont{M.~C.} \bibnamefont{Cross}} \bibnamefont{and}
  \bibinfo{author}{\bibfnamefont{H.}~\bibnamefont{Greenside}},
  \emph{\bibinfo{title}{Pattern Formation and Dynamics in Nonequilibrium
  Systems}} (\bibinfo{publisher}{Cambridge University Press},
  \bibinfo{year}{2009}), ISBN \bibinfo{isbn}{9780521770507}.

\bibitem[{\citenamefont{Blasius and T\"{o}njes}(2005)}]{Blasius05}
\bibinfo{author}{\bibfnamefont{B.}~\bibnamefont{Blasius}} \bibnamefont{and}
  \bibinfo{author}{\bibfnamefont{R.}~\bibnamefont{T\"{o}njes}},
  \bibinfo{journal}{Phys. Rev. Lett.} \textbf{\bibinfo{volume}{95}},
  \bibinfo{pages}{084101} (\bibinfo{year}{2005}).

\bibitem[{\citenamefont{Cross}(2012)}]{Cross12}
\bibinfo{author}{\bibfnamefont{M.~C.} \bibnamefont{Cross}},
  \bibinfo{journal}{Phys. Rev. E} \textbf{\bibinfo{volume}{85}},
  \bibinfo{pages}{046214} (\bibinfo{year}{2012}).

\bibitem[{\citenamefont{A.~Demir and Roychowdhury}(2000)}]{DemirMehrotra00}
\bibinfo{author}{\bibfnamefont{A.~M.} \bibnamefont{A.~Demir}} \bibnamefont{and}
  \bibinfo{author}{\bibfnamefont{J.}~\bibnamefont{Roychowdhury}},
  \bibinfo{journal}{IEEE Trans. Circuits and Syst.}
  \textbf{\bibinfo{volume}{47}}, \bibinfo{pages}{655 } (\bibinfo{year}{2000}).

\bibitem[{\citenamefont{Paullet and Ermentrout}(1994)}]{Paullet94}
\bibinfo{author}{\bibfnamefont{J.~E.} \bibnamefont{Paullet}} \bibnamefont{and}
  \bibinfo{author}{\bibfnamefont{G.~B.} \bibnamefont{Ermentrout}},
  \bibinfo{journal}{SIAM J. on Appl. Math.} \textbf{\bibinfo{volume}{54}},
  \bibinfo{pages}{1720} (\bibinfo{year}{1994}).

\bibitem[{\citenamefont{Sakaguchi}(2008)}]{Sakaguchi08}
\bibinfo{author}{\bibfnamefont{H.}~\bibnamefont{Sakaguchi}},
  \bibinfo{journal}{J. Korean Phys. Soc.} \textbf{\bibinfo{volume}{53}},
  \bibinfo{pages}{1257} (\bibinfo{year}{2008}).

\bibitem[{\citenamefont{Kuramoto}(1984)}]{Kuramotobook}
\bibinfo{author}{\bibfnamefont{Y.}~\bibnamefont{Kuramoto}},
  \emph{\bibinfo{title}{Chemical Oscillations, Waves, and Turbulence}}
  (\bibinfo{publisher}{Springer}, \bibinfo{address}{New York},
  \bibinfo{year}{1984}), ISBN \bibinfo{isbn}{9780486428819}.

\bibitem[{\citenamefont{Hagan}(1982)}]{Hagan82}
\bibinfo{author}{\bibfnamefont{P.~S.} \bibnamefont{Hagan}},
  \bibinfo{journal}{SIAM J. on Appl. Math.} \textbf{\bibinfo{volume}{42}},
  \bibinfo{pages}{762} (\bibinfo{year}{1982}).

\bibitem[{\citenamefont{T\"onjes and Blasius}(2009)}]{Tonjes09}
\bibinfo{author}{\bibfnamefont{R.}~\bibnamefont{T\"onjes}} \bibnamefont{and}
  \bibinfo{author}{\bibfnamefont{B.}~\bibnamefont{Blasius}},
  \bibinfo{journal}{Phys. Rev. E} \textbf{\bibinfo{volume}{79}},
  \bibinfo{pages}{016112} (\bibinfo{year}{2009}).

\bibitem[{\citenamefont{Cross and Hohenberg}(1993)}]{Cross93}
\bibinfo{author}{\bibfnamefont{M.~C.} \bibnamefont{Cross}} \bibnamefont{and}
  \bibinfo{author}{\bibfnamefont{P.~C.} \bibnamefont{Hohenberg}},
  \bibinfo{journal}{Rev. Mod. Phys.} \textbf{\bibinfo{volume}{65}},
  \bibinfo{pages}{851} (\bibinfo{year}{1993}).

\bibitem[{\citenamefont{Aranson and Kramer}(2002)}]{Aranson02}
\bibinfo{author}{\bibfnamefont{I.~S.} \bibnamefont{Aranson}} \bibnamefont{and}
  \bibinfo{author}{\bibfnamefont{L.}~\bibnamefont{Kramer}},
  \bibinfo{journal}{Rev. Mod. Phys.} \textbf{\bibinfo{volume}{74}},
  \bibinfo{pages}{99} (\bibinfo{year}{2002}).

\end{thebibliography}

\end{document}